\documentclass[9pt,prl,twocolumn,eqsecnum,amssymb,amsmath,a4paper,superscriptaddress,nobibnotes,floatfix]{revtex4-2}

\usepackage{graphicx}% Include figure files
\usepackage{dcolumn}% Align table columns on decimal point
\usepackage{bm}% bold math
\usepackage{hyperref}% add hypertext capabilities
\usepackage{xcolor}
\usepackage{amsmath}
\usepackage{dsfont}
\hypersetup{  colorlinks,
              linktoc         = page, % section, page, all, none
              urlcolor        = {green!80!blue},
              linkcolor       = {orange!60!black}, % footnotes+ref's
              urlcolor        = {cyan!80!black}, % reference links
              citecolor       = {cyan!80!black}, % \cite's
              anchorcolor     = {yellow}
}

\counterwithout{equation}{section}
\setcitestyle{super}% Superscript citations

\begin{document}
\title{Rotating Curved Spacetime Signatures from a Giant Quantum Vortex\vspace{-0.1cm}}

\author{Patrik \v{S}van\v{c}ara}\email{patrik.svancara@nottingham.ac.uk}
\affiliation{School of Mathematical Sciences, University of Nottingham, University Park, Nottingham, NG7 2RD, UK}
\affiliation{Nottingham Centre of Gravity, University of Nottingham,
University Park, Nottingham NG7 2RD, UK}

\author{Pietro Smaniotto}
\affiliation{School of Mathematical Sciences, University of Nottingham, University Park, Nottingham, NG7 2RD, UK}
\affiliation{Nottingham Centre of Gravity, University of Nottingham,
University Park, Nottingham NG7 2RD, UK}

\author{Leonardo Solidoro}
\affiliation{School of Mathematical Sciences, University of Nottingham, University Park, Nottingham, NG7 2RD, UK}
\affiliation{Nottingham Centre of Gravity, University of Nottingham,
University Park, Nottingham NG7 2RD, UK}

\author{James F. MacDonald}
\affiliation{School of Physics and Astronomy, University of Nottingham, University Park, Nottingham, NG7 2RD, UK}

\author{Sam~Patrick}
\affiliation{Department of Physics, King’s College London, University of London, Strand, London, WC2R 2LS, UK}

\author{Ruth Gregory}
\affiliation{Department of Physics, King’s College London, University of London, Strand, London, WC2R 2LS, UK}
\affiliation{Perimeter Institute, 31 Caroline Street North, Waterloo, ON, N2L 2Y5, Canada}

\author{Carlo F. Barenghi}
\affiliation{School of Mathematics, Statistics and Physics, Newcastle University, Newcastle upon Tyne, NE1 7RU, UK}

\author{Silke Weinfurtner}
\email{silke.weinfurtner@nottingham.ac.uk}
\affiliation{School of Mathematical Sciences, University of Nottingham, University Park, Nottingham, NG7 2RD, UK}
\affiliation{Nottingham Centre of Gravity, University of Nottingham,
University Park, Nottingham NG7 2RD, UK}
\affiliation{Perimeter Institute, 31 Caroline Street North, Waterloo, ON, N2L 2Y5, Canada}
\affiliation{Centre for the Mathematics and Theoretical Physics of Quantum
Non-Equilibrium Systems, University of Nottingham, University Park, Nottingham, NG7 2RD, UK}

\maketitle
\onecolumngrid
\vspace{-0.6cm}

\par\noindent
Gravity simulators \cite{barcelo2011} are laboratory systems where small excitations like sound~\cite{unruh1981} or surface waves~\cite{schutzhold2002,barroso2023} behave as fields propagating on a curved spacetime geometry. The analogy between gravity and fluids requires vanishing viscosity~\cite{unruh1981,schutzhold2002,barroso2023}, a feature naturally realised in superfluids like liquid helium or cold atomic clouds~\cite{clovecko2019,Lahav2010,viermann2022,tajik2023}. Such systems have been successful in verifying key predictions of quantum field theory in curved spacetime~\cite{viermann2022,tajik2023,steinhauer2014,steinhauer2016,braidotti2022}. In particular, quantum simulations of rotating curved spacetimes indicative of astrophysical black holes require the realisation of an extensive vortex flow~\cite{visser2005} in superfluid systems. Here we demonstrate that despite the inherent instability of multiply quantised vortices~\cite{shin2004,patrick2022}, a stationary giant quantum vortex can be stabilised in superfluid $^4$He. Its compact core carries thousands of circulation quanta, prevailing over current limitations in other physical systems such as magnons~\cite{clovecko2019}, atomic clouds~\cite{Lahav2010, viermann2022} and polaritons~\cite{cookson2021,alperin2021}. We introduce a minimally invasive way to characterise the vortex flow~\cite{moisy2009,wildeman2018} by exploiting the interaction of micrometre-scale waves on the superfluid interface with the background velocity field. Intricate wave-vortex interactions, including the detection of bound states and distinctive analogue black hole ringdown signatures, have been observed. These results open new avenues to explore quantum-to-classical vortex transitions and utilise superfluid helium as a finite temperature quantum field theory simulator for rotating curved spacetimes~\cite{braunstein2023}.
\bigskip

\twocolumngrid
To experimentally realise a curved spacetime such as a black hole requires a specific relative motion between the excitations and the background medium. One-dimensional supersonic flow, the archetypal example of an acoustic black hole, provides a platform for observations of Hawking radiation in both classical~\cite{weinfurtner2011,euve2016} and quantum fluids~\cite{steinhauer2014,steinhauer2016,jacquet2023}. More complex phenomena like Penrose superradiance require rotating geometries realisable in two spatial dimensions, e.g. by means of a stationary draining vortex flow~\cite{visser2005,solnyshkov2019}. Classical fluid flow experiments have demonstrated the power of the gravity simulator program, realising superradiant amplification of both coherent~\cite{torres2017,braidotti2022} and evanescent waves~\cite{cromb2020}, as well as quasinormal mode oscillations~\cite{torres2020}, a process intimately connected to black hole ringdown~\cite{abbott2016}.

Here, we investigate related phenomena in the limit of negligible viscosity in superfluid $^4$He (called He~II). Its energy dissipation is temperature dependent and can be finely adjusted across a wide range. At $1.95$~K, where our experiments take place, its kinematic viscosity is reduced by a factor of $100$ compared to water~\cite{donnelly1998} and the damping is dominated by thermal excitations collectively described by the viscous normal component~\cite{barenghi2014intro,donnelly1998} that constitutes approximately half of the total density of the liquid. Moreover, He~II supports the existence of line-like topological defects called quantum vortices. Each vortex carries a single circulation quantum $\kappa \approx 10^{-7}$~m$^2$/s and forms an irrotational (zero-curl) flow field in its vicinity~\cite{barenghi2014intro}. Due to this discretisation, a draining vortex of He~II can manifest itself only as a multiply quantised (also known as giant) vortex or as a cluster of single quantum vortices. Such vortex bundles exhibit their own collective dynamics and can even introduce solid body rotation~\cite{feynman1957} at length scales larger than the inter-vortex distance, adding complexity to the study of quantum fluid behaviour. As the realisation of curved spacetime scenarios requires an irrotational velocity field~\cite{barcelo2011,fischer2002}, it is critical to confine any rotational elements into a central area, i.e. the vortex core. However, alike-oriented vortices have a tendency to move apart from each other, which poses a limitation on the extent of the core one can stabilise in an experiment. On the other hand, recent findings show that mutual friction~\cite{barenghi2014intro} between quantum vortices and the normal component contributes to the stabilisation of dense vortex clusters~\cite{galantucci2023}.

The vortex induces a specific velocity field within the superfluid, which affects the propagation of small waves on its surface. In particular, low frequency excitations perceive an effective acoustic metric~\cite{schutzhold2002,barroso2023}
\begin{equation}
      g_{ij} \propto \begin{pmatrix}-c^2+v^2 & -\bm{v} \\ -\bm{v} & \mathds{1}_{2\times 2}\end{pmatrix},
      \label{eq:metric}
\end{equation}
where $c$ denotes their propagation speed and $\bm{v}(r,\theta) = v_r \hat{\bm{r}} + v_\theta \hat{\bm{\theta}}$ indicates the velocity field at the interface (we assume that the superfluid and normal velocity fields are equal, in line with other mechanically driven flows of He~II~\cite{varga2018,diribarne2021}). Although this description fails in the high frequency regime due to dispersion, it is well known that the curved spacetime phenomenology persists for these excitations~\cite{torres2017,torres2020,patrick2020}. Altogether, the above properties suggest that an extensive draining vortex of He~II is a feasible candidate for simulations of a quantum field theory in curved spacetime.

\begin{figure}[h!]
  \centering
  \includegraphics[width=\columnwidth]{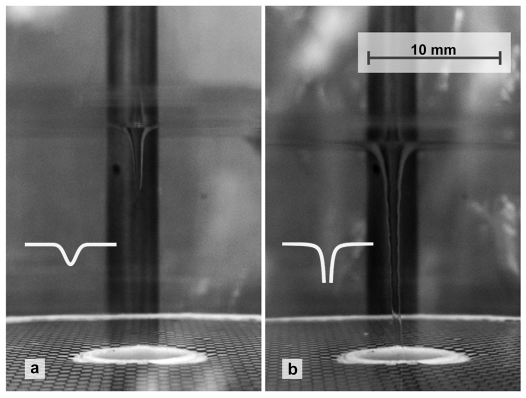}
  \caption{\textbf{Side views of two distinct configurations of the giant quantum vortex}. \textbf{a,} At low propeller frequencies (here, $1$~Hz), the interface exhibits a discernible depression while the vortex core beneath takes the form of a compact, polarised cluster of singly quantised vortices (called solid core). \textbf{b,} With the escalation of frequency (here, to $2$~Hz), a fully formed hollow core emerges, behaving like a multiply quantised object. Dark vertical stripes in the background provide contrast to the imaged interface. A simplified sketch of said interface (white lines) helps to identify these regimes in later figures.}
  \label{fig:setup}
\end{figure}

We realised this flow in cylindrical geometry that is built upon the concept of a stationary suction vortex~\cite{yano2018} (see Methods for a detailed description). The central component of our set-up is a spinning propeller, which is responsible for establishing a continuous circulating loop of He~II, feeding a draining vortex that forms in the optically accessible experimental zone. At small propeller speeds, we observe a depression on the superfluid interface ({Fig.~\ref{fig:setup}a}), but as the speed increases, this depression deepens and eventually transforms into a hollow vortex core extending from the free surface to the bottom drain ({Fig.~\ref{fig:setup}b}). The parabolic shape of the free surface in the former regime is consistent with solid body rotation, which corresponds to a compact, polarised cluster of singly quantised vortices (called solid core) that forms under the finite depression. The hollow core can instead absorb individual circulation quanta and behave like a multiply quantised object~\cite{inui2020}. To minimise the rotational flow injected by the spinning propeller into the experimental zone, we devised a unique recirculation strategy based on a purpose-built flow conditioner (see Methods) that promotes formation of a centrally confined vortex cluster instead of a sparse vortex lattice. However, the exact dynamics of individual quantum vortices, as well as their spatial distribution in the experiment, calls for future investigations. State-of-the-art numerical models~\cite{galantucci2020} account for the motion of vortex lines coupled to the superfluid and normal velocity fields, but fail to dynamically model the interface, which is a pivotal element in our system. Previous experimental efforts~\cite{matsumura2019,obara2021,kakimoto2022} confirmed that a draining vortex in He~II carries macroscopic circulation, but lacked spatial resolution required to investigate central confinement of rotational components. In this regard, cryogenic flow visualisation~\cite{guo2014} provides sufficient resolution. However, this method requires probes in the form of small solid particles into the superfluid, which accumulate along the vortex lines and considerably affect their dynamics~\cite{sergeev2009}.

The above limitations compelled us to propose an alternative, minimally invasive method to probe the vortex flow and extract macroscopic flow parameters that exploits the relative motion occurring between interface waves and the underlying velocity field. The corresponding dispersion relation for angular frequencies $\omega$ and wave vectors $\bm{k}$ reads~\cite{patrick2020} 
\begin{equation}
    \label{eq:dispersion}
    (\omega - \bm{v}\cdot\bm{k})^2 = F(||\bm{k}||)\,,
\end{equation}
where $F$ denotes the dispersion function. By solving Eq.~\eqref{eq:dispersion}, we find (see Methods) that the spectrum of interface modes gets frequency shifted and the velocity field can be inferred from these shifts~\cite{liu2021}. Therefore, we redirect our attention towards precise detection of small waves propagating on the superfluid interface.

\begin{figure*}
  \centering
  \includegraphics[width=\linewidth]{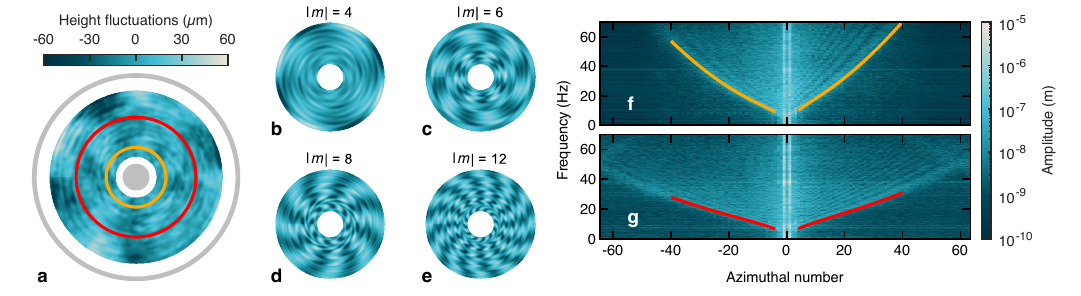}
  \caption{\textbf{Superfluid interface reconstruction and wave analysis}. \textbf{a,} Snapshot of the free helium surface depicts height fluctuations representing micrometre waves excited on the superfluid interface. Grey areas mark the positions of the central drain (radius $5$~mm) and the outer glass wall (radius $37.3$~mm). \textbf{b-e,} Examples of different azimuthal modes $|m|$ ($m$ counting the number of wave crests or troughs around a circular path) extracted from panel \textbf{a} by a discrete Fourier transform. Wave amplitudes are rescaled for better visibility. \textbf{f-g,} Two-dimensional wave spectra obtained by transforming angle and time coordinates, for radii of $11.2$ (panel \textbf{f}) and $22.1$~mm (panel \textbf{g}). These radii are marked in panel \textbf{a} by coloured circles. Absence of excitations in low frequency bands (below coloured lines) can be understood through the solution of Eq.~\eqref{eq:dispersion}. The corresponding theoretical predictions of the minimum frequency permissible for propagation for the given radii can be matched with experimental observations (yellow and red lines).}
  \label{fig:reconstruction}
\end{figure*}

We identified that the adapted Fourier transform profilometry~\cite{moisy2009,wildeman2018} is well suited to our needs, as it is capable of resolving a fluid interface with sufficient and simultaneous resolution in both space and time. This powerful technique consists in imaging the disturbed interface against a periodic backdrop pattern. This way, we resolve height fluctuations of said interface ({Fig.~\ref{fig:reconstruction}a}) with sensitivity up to approximately one micrometre. Due to symmetries of the flow, the waves exhibit two conserved quantities: frequency $f$ and azimuthal number $m$. The latter parameter counts the number of wave crests around a circular path, with positive or negative values of $m$ corresponding to wave patterns co- or counter-rotating with the central vortex.

These spatial patterns (or modes) can be retrieved from the height fluctuation field by a discrete Fourier transform. For example, by transforming with respect to the angle $\theta$ one can single out individual azimuthal modes ({Fig.~\ref{fig:reconstruction}b-e}). To study wave dynamics in time, we must also transform the temporal coordinate and inspect the resulting two-dimensional spectra, showcased in {Fig.~\ref{fig:reconstruction}f-g} for two distinct radii. Notable high-amplitude signals in the $m=\pm 1$ bands are exclusively a consequence of how mechanical vibrations of the set-up imprint themselves upon our detection method. Of physical interest are modes with higher azimuthal numbers. These excitations, observed in both solid and hollow core regimes, represent micrometre waves excited on the interface. In the steady state, the waves dissipate their energy, in part by viscous damping and in part by scattering into the draining core of the vortex~\cite{patrick2021}. Although this is balanced by the stochastic drive originating from the fluid flow and/or aforementioned mechanical vibrations, we notice that only a certain region of the spectral space $(m,f)$ is populated with excitations, a feature that varies when examining smaller ({Fig.~\ref{fig:reconstruction}f}) and larger radii ({Fig.~\ref{fig:reconstruction}g}). We observe that only some high-frequency (equivalent to high-energy) waves have the capability to propagate on the interface. Through the solution of Eq.~\eqref{eq:dispersion}, we can pinpoint the minimum frequency, $f_{\min}$ permissible for propagation for the given radius, azimuthal number and background velocity (see Methods) and, in line with the methodology introduced above, we exploit this particular frequency to extract the underlying velocity field as we now describe. We search the parameter space produced by two velocity components $(v_r,v_\theta)$ and determine values that produce the best match between $f_{\min}$ and the lowest excited frequency in the experimental data across multiple azimuthal modes (coloured lines in {Fig.~\ref{fig:reconstruction}f-g}). By carrying out this procedure for every examined radius, we can reconstruct the velocity distribution in the draining vortex flow.
\newpage

\begin{figure}[h!]
  \centering
  \includegraphics[width=\columnwidth]{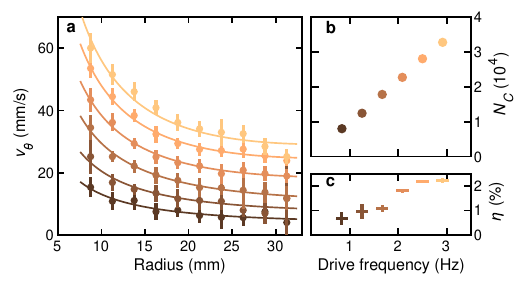}
  \caption{\textbf{Reconstructed velocity distribution and flow parameters}. \textbf{a,} Coloured points denote the radial dependence of the azimuthal velocity $v_\theta$ for six vortex configurations distinguished by the drive (propeller) frequency. Each point is obtained by averaging over a $2.5$~mm radial interval. Radial velocity component is approximately zero across all instances. Best fits of $v_\theta(r)$ (coloured lines) yield the circulation $C$ of the central vortex and the angular frequency $\Omega$ of the additional solid body rotation. \textbf{b,} Number of circulation quanta confined in the vortex core, $N_C = 2\pi C/\kappa$ corresponds to the most extensive vortex structures ever observed in quantum fluids. Vertical error bars are comparable to the symbol size. \textbf{c,} The ratio $\eta$ between $\Omega$ and the angular frequency of the drive is less than $2.5\%$ in all cases, suggesting that the velocity field in our system is dominated by the irrotational vortex flow.}
  \label{fig:velocity}
\end{figure}

We conducted these reconstructions across multiple vortex configurations distinguished by the drive (propeller) frequency. For all instances, $v_r$ approximates zero within the limits of our resolution. Although seemingly paradoxical, this outcome results from a complex boundary layer interaction and is in consonance with earlier findings in classical fluids~\cite{andersen2003}. Therefore, interface waves engage with an almost entirely circulating flow characterised by a specific radial dependence of $v_\theta$ (coloured points in {Fig.~\ref{fig:velocity}a}). Overall, the results are consistent with
\begin{equation}
    \label{eq:velocity}
    v_{\theta} (r) = \Omega r + \frac{C}{r},
\end{equation}
indicated in {Fig.~\ref{fig:velocity}a} by coloured lines. The first term represents solid body rotation with angular frequency $\Omega$, which leaks into the experimental area through the flow conditioner as described above. The second term corresponds to an irrotational flow around a central vortex with circulation $C$. The related number of circulation quanta confined in its core, $N_C = 2\pi C /\kappa$ is displayed in {Fig.~\ref{fig:velocity}b} as a function of the drive frequency. Across all instances, the core consists of the order of $10^4$ quanta, a record-breaking value in the realm of quantum fluids. In the solid core regime, $N_C$ can be identified with the number of individual quantum vortices concentrated in the core. However, in the context of a hollow core, $N_C$ represents its topological charge. Achieving circulation values separated from the elementary quantum $\kappa$ by four orders of magnitude allows the quantisation of circulation to be disregarded, rendering the vortex effectively classical. This unprecedented realisation of a giant quantum vortex flow represents a distinctive instance of a quantum-to-classical flow transition in He~II~\cite{lamantia2014}.

Significance of the aforementioned outcomes can be underlined by noting that $n$-quantised vortices are dynamically unstable~\cite{shin2004,patrick2022}. They spontaneously decay into a cluster of $n$ vortices~\cite{geelmuyden2022} due to the excitation of a negative energy mode in the multiply-quantised vortex core~\cite{braidotti2022,geelmuyden2022}. Nevertheless, dynamical stabilisation of giant vortices can be achieved by suitably manipulating the superfluid. Namely, introducing a draining flow and reducing the fluid density at the centre has proven effective in polariton condensates, for vortices with $n \lesssim 100$ (ref.~\cite{cookson2021,alperin2021}). These results align harmoniously with our experiment, where the reduced density translates into the existence of a hollow core and the draining flow resides in the bulk of the draining vortex.

It is worth noting that larger circulation values around a draining vortex in He~II are documented in literature~\cite{kakimoto2022}. However, therein the contributions of the vortex core and the solid body rotation are not distinguished. The second effect may dominate in the reported circulations, since the number of quantum vortices responsible for rotation~\cite{feynman1957} scales with the corresponding angular frequency $\Omega$. Rotation in our experimental zone is notably suppressed. The sparse presence of quantum vortices partially justifies our assumption that normal and superfluid components behave as a single fluid. More importantly, the ratio $\eta$ between $\Omega$ and the angular frequency of the drive does not exceed $2.5$\% ({Fig.~\ref{fig:velocity}c}), and the velocity field in our system is dominated by the irrotational vortex flow. The core of this vortex must be smaller than $7.6$~mm, the smallest probed radius, because the velocity profiles ({Fig.~\ref{fig:velocity}a}) show no indication of a turning point at small radii.

\begin{figure*}
  \centering
  \includegraphics{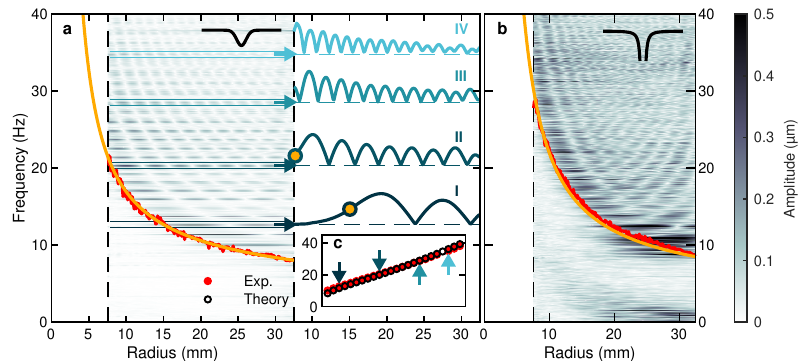}
  \caption{\textbf{Bound states in co-rotating waves}. Fourier amplitudes of interface waves corresponding to $m = 8$ mode reveal a characteristic pattern in the radial direction that can be identified with bound states, i.e. standing waves between the outer boundary (glass wall) at $37.3$~mm and the effective potential barrier (red lines). A simplified, but accurate model of the potential (yellow lines) is extended beyond the experimentally accessible range (dashed black lines). \textbf{a,} Solid core regime. Rescaled amplitudes of four bound states labelled I-IV (blue lines) are shown as a function of radius. Crossing points with the potential barrier are marked by yellow points. \textbf{b,} Hollow core regime. \textbf{c,} Comparison of bound state frequencies retrieved from panel \textbf{a} (red points) and their theoretical predictions (black circles). Frequencies of states I-IV are highlighted by blue arrows.}
  \label{fig:cobound}
\end{figure*}

We can nonetheless venture beyond the experimental range by exploring wave dynamics in the radial direction. We restrict our discussion to a particular mode $|m| = 8$ ({Fig.~\ref{fig:reconstruction}d}) as a representative of the outlined behaviour. We start by analysing co-rotating ($m = 8$) modes, displayed in {Fig.~\ref{fig:cobound}a-b} for the solid and hollow core structures. In both cases, $f_{\min}$ (red line) denotes an effective potential barrier, preventing waves from reaching the vortex core. Existence of this barrier, together with an outer, solid boundary at $37.3$~mm, gives rise to bound states (standing waves), appearing as distinct, striped patterns extending up to $40$~Hz. These patterns represent the first direct measurement of resonant surface modes around a macroscopic vortex flow in He~II.

To perform an in-depth examination of selected states (denoted as {I-IV}), we plot the absolute value of their amplitudes in {Fig.~\ref{fig:cobound}a}. The frequency of state {I} meets $f_{\min}$ in a crossing point (yellow point) located within the field of view. At large radii, this wave harmonically propagates. However, as it penetrates the barrier, its amplitude exponentially decays in exact analogy with a simple quantum-mechanical model of a particle trapped in a potential well. Upon escalating the frequency, the crossing point moves towards smaller radii (state {II}), eventually reaching the limit of our detection range. For high-frequency states ({III} and {IV}), the crossing point is well outside the detection range and we only observe the harmonic part of the signal. Nonetheless, the mere existence and predictability of these states lets us extend the effective potential barrier beyond the observable range.

Specifically, we consider a model of a purely circulating vortex, whose velocity field reads $(v_r, v_\theta) = (0, C/r)$ and extend the experimentally determined potential barrier (red lines in {Fig.~\ref{fig:cobound}a-b}) towards smaller radii (yellow lines). In practice, this model must breakdown near the vortex core, where the spatial distribution of individual quantum vortices becomes relevant. Nonetheless, the frequencies of individual bound states are in excellent agreement with theoretical predictions (see Methods) based on the extended potential barrier ({Fig.~\ref{fig:cobound}c}). This outcome validates the simplified model and allows us to constrain the radius of the core region to approximately $4$ and $6$~mm, respectively for the solid and hollow core regimes. Confinement of the rotating core beyond the experimental range gains significance when considering the draining vortex flow as a gravity simulator, for example when searching for initial indications of black hole ringdown.

\begin{figure*}
\centering
  \parbox[t][][b]{\columnwidth}{\includegraphics[width=\columnwidth]{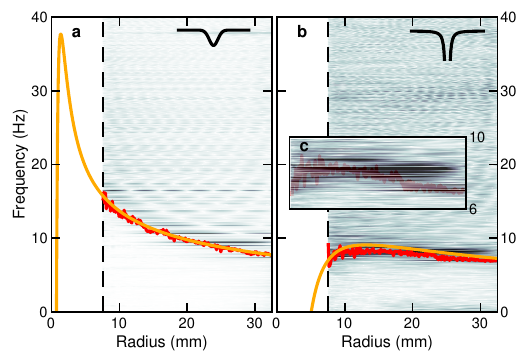}}\hfill%
  \parbox[t][][t]{\columnwidth}{\caption{\textbf{Bound states and ringdown modes in counter-rotating waves}. Fourier amplitudes of interface waves (same colour scale as in Fig.~\ref{fig:cobound}) corresponding to $m =-8$ mode interact with the effective potential barrier (red lines). Its simplified model (yellow lines) is extended beyond the accessible range (dashed black lines). \textbf{a,} In the solid core regime, the potential allows existence of bound states, visible up to approximately $30$~Hz. \textbf{b,} In the hollow core regime, no bound states can be retrieved. Instead, we observe dominant excitations lingering near the shallow maximum of the potential (approximately at $8.25$~Hz), suggesting the excitation of black hole ringdown modes. \textbf{c,} Inset highlights ringdown mode candidates from panel \textbf{b}, with the effective potential barrier displayed as a pale red line.}
  \label{fig:counterbound}}
  \vspace{-1.5em}
\end{figure*}

For this purpose, we focus on counter-rotating ($m=-8$) modes, depicted in {Fig.~\ref{fig:counterbound}a-b} with the effective potential barriers (red lines) and their extensions (yellow lines). The shape of the barrier in the solid core regime ({Fig.~\ref{fig:counterbound}a}) allows the existence of bound states up to approximately $30$~Hz. However, this is not the case in the hollow core regime ({Fig.~\ref{fig:counterbound}b}) despite the corresponding circulations only differing within one order of magnitude. Bound states are not formed at all since the effective potential displays a shallow maximum before decreasing towards zero. Dominant excitations in this spectrum, highlighted in {Fig.~\ref{fig:counterbound}c}, are modes lingering near this maximum. These excitations, previously identified as ringdown modes of an analogue black hole~\cite{torres2020}, represent the very first hints of this process taking place in a quantum fluid. The radius where the effective potential crosses the zero frequency level is related to the analogue ergoregion~\cite{patrick2020}, a key feature in the occurrence of black hole superradiance. To directly observe this region in our set-up, further increasing the azimuthal velocity and/or probing the system closer to the vortex core is required.
\newpage

Our research positions quantum liquids, particularly He~II, as promising contenders for finite temperature, non-equilibrium quantum field theory simulations, marking a transformative shift from already established simulators in curved spacetimes~\cite{viermann2022,tajik2023,steinhauer2014,steinhauer2016}. The liquid nature of He~II arises from an effective, strongly interacting field that complements its weakly interacting counterpart found, e.g. in cold atomic clouds. A distinctive advantage presented by He~II lies in its flexibility, allowing it to be operated at a fixed temperature, starting just below the superfluid transition where He~II displays pronounced dissipation. This regime in particular holds immense potential, e.g. for the mapping to generic holographic theories~\cite{wittmer2021}. At temperatures below $1$~K, the normal component is expected to be an aggregate of individual thermal excitations. This tunability provides the opportunity to investigate a broad spectrum of finite temperature quantum field theories.

Owing to the capacity of He~II to accommodate macroscopic systems, we achieved the creation of extensive vortex flows in a quantum fluid. Notably, the hollow vortex core's size scales with its winding number, and consequently, system size constraints may restrict the maximum circulation achievable when implemented in cold atom or polariton systems alike. Key processes in rotating curved spacetimes, such as superradiance and black hole ringing, can be explored in our current system with minor adjustments to the propeller speed, container geometry, or by dynamically varying flow parameters. Our set-up also provides a distinctive opportunity to investigate rotating curved spacetimes with tunable and genuinely quantised angular momentum, setting it apart from classical liquids. Furthermore, applying these techniques to explicitly time-dependent scenarios allows for the exploration of fundamental non-equilibrium field theory processes. This may involve controlled modulations of first or second sound in the bulk of the quantum liquid, providing a platform for conducting wave turbulence simulations across various length and temperature scales. This represents a noteworthy advancement beyond the current scope of cold atom studies~\cite{dogra2023}.

\medskip\textbf{Acknowledgements---}%
We deeply appreciate the technical team at the School of Physics and Astronomy, University of Nottingham, for their unwavering commitment during the Covid-19 pandemic. Their efforts led to the establishment of a new cryogenic laboratory. Special thanks to Terry Wright for his expertise in realising this vision. We are grateful to John Owers-Bradley for his crucial support in setup planning and design. Viktor Tsepelin's contributions in heat leak mitigation were indispensable. We also thank Xavier Rojas and Gr\'{e}goire Ithier for discussions on light integration into the cryogenic system. Acknowledgements to all members of the Gravity Laboratory group at the University of Nottingham for their insightful discussions and contributions, particularly Vitor Barroso Silveira, whose work with Fourier transform profilometry greatly enriched our project. Furthermore, S.W., R.G., C.F.B., and P.\v{S}. extend their appreciation to the Science and Technology Facilities Council for their generous support in Quantum Simulators for Fundamental Physics (QSimFP), (ST/T006900/1, ST/T005858/1, and ST/T00584X/1), as part of the UKRI Quantum Technologies for Fundamental Physics program. S.W., L.S. and J.F.M. gratefully acknowledge the  support of the Leverhulme Research Leadership Award (RL-2019-020). S.W. also acknowledges the Royal Society University Research Fellowship (UF120112). R.G. and S.W. expresses sincere appreciation to the Perimeter Institute for Theoretical Physics for their warm hospitality and organisation of the QSimFP network meeting. Research conducted at the Perimeter Institute is made possible through the generous support of the Government of Canada, via the Department of Innovation, Science, and Economic Development Canada, as well as by the Province of Ontario, through the Ministry of Colleges and Universities.

\medskip\textbf{Author contributions---}%
All authors contributed substantially to the work. Overall conceptualisation by S.W. Further development of experimental design by P.\v{S}, J.F.M., and S.W. Data collection by P.\v{S}, P.S., and L.S. Analysis: Theoretical aspects of wave-current interaction techniques developed by S.P., S.W., and R.G. Interface detection method development by P.\v{S}, P.S., L.S., and S.W. Manuscript writing by P.\v{S}, P.S., L.S., S.P., and S.W. Review and editing by R.G. and C.F.B. Funding acquisition by S.W., R.G., and C.F.B. Overall project supervision by S.W. Theory coordination by R.G. and C.F.B. PhD student co-supervision by P.\v{S}. S.W. is the corresponding author.

\medskip\textbf{Data and code availability---}%
The datasets generated and analysed during the current study are available from the corresponding author. The conclusions of this study do not depend on code or algorithms beyond standard numerical evaluations.

\medskip\textbf{Competing interests---}%
The authors declare no competing interests.

\medskip\textbf{Materials \& Correspondence---}%
    Correspondence should be addressed to S.W. and requests for materials should be addressed to P.\v{S}.

\interlinepenalty=10000
\bibliography{biblio}

\clearpage \raggedbottom

\interlinepenalty=0
\section{Methods}

\subsection{Cryogenic drive}

\begin{figure}[h!]
  \centering
  \includegraphics[width=\columnwidth]{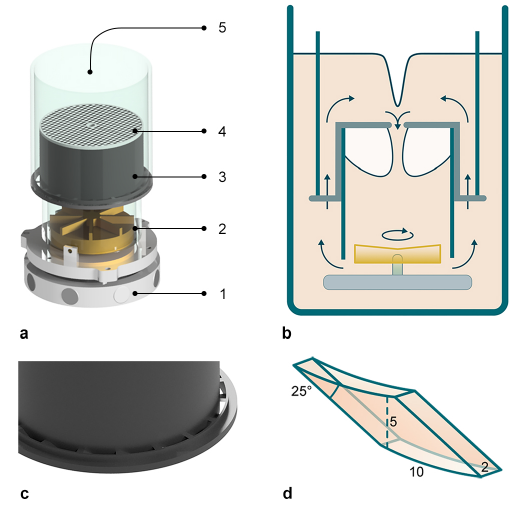}
  \caption{\textbf{Cryogenic realisation of a stationary giant quantum vortex.} \textbf{a,} A disk (1) is magnetically coupled to a room temperature drive to achieve rotation frequencies of up to $3$~Hz. The disk drives a propeller (2) functioning as a centrifugal pump, establishing a continuous circulation loop of He~II through a purpose-built flow conditioner (3) decorated with a periodic pattern (4). A draining vortex forms in the centre of the designated experimental zone (5). \textbf{b,} The schematic illustrates the outline of our recirculation technique. \textbf{c,} Detail on the edge of the flow conditioner equipped with 18 slanted openings. \textbf{d,} Schematic representation of one such opening with dimensions given in millimetres.}
  \label{fig:setup2}
\end{figure}

The experimental set-up (Fig.~\ref{fig:setup2}a) relies on the use of a spinning propeller that acts as a centrifugal pump, establishing a steady recirculation loop of He~II between the experimental zone and the area underneath (Fig.~\ref{fig:setup2}b). The draining vortex forms above a circular aperture, $10$~mm in diameter and located at the top of a purpose-built flow conditioner (Fig.~\ref{fig:setup2}c), whose geometry was optimised through experimentation. Namely, the flow conditioner features 18 slanted openings (Fig.~\ref{fig:setup2}d) regularly placed along its outer edge, which are used to reinject He~II into the experimental zone. In order to suppress the amount of undesired solid body rotation that leaks into the experimental zone from the vicinity of the propeller, we adopt the strategy of driving the propeller in the direction opposite to the orientation of these openings. We eventually find that the direction of the draining vortex that forms in the experimental zone coincides with the direction of these openings.

\medskip
In order to establish a stable recirculation of He~II, it is crucial to rely on a constant drive of the propeller. For this purpose, a robust connection between the propeller submerged in He~II and its room temperature drive is achieved by means of magnetic coupling. In particular, a magnetic ring rotates outside the cryostat driven by a DC motor. Inside the cryostat, a solid disk decorated with permanent magnets (see again Fig.~\ref{fig:setup2}a (1)) ensures that motion is transferred to the propeller (Fig.~\ref{fig:setup2}a (2)). This way, we achieve a smooth operation of the drive, with steadfast connection up to approximately $3$~Hz, ensuring continuous and reliable power transmission to the low temperature environment. The need for magnetic coupling led us to enclose the set-up in a custom cryogenic system made of glass. Its fully transparent design provides excellent optical access, while maintaining a constant temperature of the helium bath within the range down to $1.55$~K and with stability better than $1$~mK.

\subsection{Interface wave dynamics}

We consider a potential flow $\bm{v} = \nabla\Phi$ and a free surface located at $z=h_0$. Perturbations of the velocity potential and free surface are denoted $\delta\Phi$ and $\delta h$, respectively. The equations of motion for surface waves are obtained by linearising the Bernoulli equation and integrating Laplace's equation from the hard boundary at $z=0$ to the free surface,
\begin{equation} \label{eq:wave-eq}
\begin{split}
    (\partial_t + \bm{v}\cdot\nabla)\delta\Phi + g\delta h - \frac{\sigma}{\rho}\nabla^2\delta h & = 0, \\
    (\partial_t + \nabla\cdot\bm{v})\delta h + i\nabla\cdot\tanh(-ih_0\nabla)\delta\Phi & = 0,
\end{split}
\end{equation}
where $g$ is the gravitational acceleration, $\sigma = 3.06 \times 10^{-4}$~N/m is the surface tension and $\rho = 145.5$~kg/m$^3$ is the density of the fluid (numerical values are given for He~II at $1.95$~K~\cite{donnelly1998}). These equations can be derived from the following action,
\begin{equation}
\begin{split}
    S[\delta\Phi,\delta h] = & \int dt \, d^2\bm{x}\bigg[-\delta h\left(\partial_t+\bm{v}\cdot\nabla\right)\delta\Phi - \frac{1}{2}g\delta h^2 \\
    & \qquad - \frac{\sigma}{2\rho}\left\vert\nabla\delta h\right\vert^2 - \frac{1}{2}\delta\Phi \mathcal{D}\left(-i\nabla\right)\delta\Phi\bigg],
\end{split}
\end{equation}
where $\mathcal{D}(\bm{k}) = k\tanh(h_0k)$ and $k=||\bm{k}||$. This constitutes the effective field theory of our system once the bulk degrees of freedom have been integrated out. For $h_0k\ll 1$, the effect of the inhomogeneous fluid flow is completely encapsulated by the effective metric given in Eq.~\eqref{eq:metric}~\cite{barcelo2011}. Wavelengths in our system satisfy the opposite limit, $h_0k\gg 1$. In this regime, solution of Eq.~\eqref{eq:wave-eq} can be achieved using the Wentzel-Kramers-Brillouin (WKB) approximation, where the local wavelength is assumed much shorter than the scale over which the background flow varies.

Within the WKB approximation, the solution of Eq.~\eqref{eq:wave-eq} is modelled as a plane wave with an amplitude and wave vector that vary smoothly as we move through the system. Assuming that $\bm{v}$ only depends on the radial coordinate and recalling that the angular frequency $\omega$ and the azimuthal number $m$ are conserved quantities within our system, we can write
\begin{equation}
    \delta\Phi\sim \exp \left( i\int p(r) dr + im\theta - i\omega t \right),
\end{equation}
where $p(r)$ is the radial part of the wave vector $\bm{k} = p(r)\hat{\bm{r}} + (m/r) \hat{\bm{\theta}}$. The amplitude is similarly only a function of $r$. At leading order, the WKB method gives the local dispersion relation~\eqref{eq:dispersion} with the dispersion function,
\begin{equation}
    F(k) = \left( gk + \frac{\sigma}{\rho} k^3 \right)\tanh(h_0 k),
\end{equation}
which is only a function of the modulus of the wave vector $k$.

As discussed in the main text, the background flow can be accurately modelled as $\bm{v} = (C/r)\hat{\bm{\theta}}$. By further assuming that $h_0$ is constant in the window of analysis, the solutions of Eq.~\eqref{eq:dispersion} define two branches of the dispersion relation,
\begin{equation} \label{eq:branches}
    \omega^\pm_D = \frac{mC}{r^2} \pm \sqrt{F(k)}.
\end{equation}
We are interested in the behaviour of Eq.~(\ref{eq:branches}) as a function of the azimuthal number $m$. 
For $m \neq 0$, we can distinguish two different contributions, one explicitly appears as $mC/r^2$ while the second is hidden in the modulus of $\bm{k}$, which is proportional to $m^2/r^2$. The effect of the latter is to widen the gap between $\omega_D^+$ and $\omega_D^-$ when approaching the vortex core, i.e. when decreasing $r$. The $mC/r^2$ term shifts both branches up for $m>0$ or down for $m<0$. For large enough $||\bm{v}||$, it is possible for $\omega_D^-$ to cross into the upper half plane ($\omega>0$) for $m>0$ or for $\omega_D^+$ to cross into the lower half plane ($\omega<0$) for $m<0$. However, this does not occur in our window of analysis, which excludes the vortex core, and therefore $\mathrm{sgn}(\omega_D^\pm)=\pm 1$ for all $r$ and $p$ considered. From here on, we restrict our attention to positive frequency modes, for which only $\omega_D^+$ is relevant.

For a given angular frequency $\omega$, the solutions of the dispersion relation $\omega=\omega_D^+(m,r,p)$ at a particular radius determine the allowed values of the radial wave vector $p$. There are a maximum of two solutions with $p\in\mathbb{R}$ for the considered flow profile, which correspond to the (radially) in-going and out-going waves far away from the vortex core. The function $\omega_D^+$ presents a minimum frequency $\omega_{\min}$ below which there are no intersections between a line of constant $\omega$ with $\omega_D^+$ at any real $p$. Therefore, $\omega_{\min}$ sets the minimum frequency for a wave to be able to propagate in the system at that particular radius. We can find this frequency by solving $\partial_p\omega_D^+ = 0$, which gives $p=0$. By inserting the solution into the dispersion relation we obtain $\omega_{\min}(m,r) = \omega_D^+(m,r,p=0)$, which (after dividing by $2\pi$) gives the yellow lines displayed on Figs.~\ref{fig:cobound}a-b and~\ref{fig:counterbound}a-b.

For $\omega < \omega_{\min}$, $p\in\mathbb{C}$ and the solutions become evanescent. In line with the main text, $\omega_{\min}$ doubles as an effective potential barrier, trapping any propagating solutions in the radial direction between the crossing point $r_\mathrm{tp}$ (defined by $\omega = \omega_{\min}(r_\mathrm{tp})$) and the outer boundary at $r_B$. The crossing point can be also understood as the turning point of the dispersion relation, i.e. the location where an in-going wave comes to rest (as seen from the vanishing of the radial part of the group velocity, $\partial_p\omega_D^+(r_\mathrm{tp}) = 0$) and gets reflected. When this happens, the two roots $p^\pm$ of the positive form of Eq.~\eqref{eq:branches} become degenerate and there is a divergence in the WKB amplitude, causing this approximation to breakdown. However, there is a known procedure~\cite{patrick2020} to circumvent this pathological behaviour as we now describe.

The WKB solution in the vicinity of an arbitrary point $r_j$ may be written as
\begin{multline}
    \delta\Phi \sim \alpha_j^+ \exp\left({i\int^{r}_{r_j} p(r') dr'}\right) + \\
        + \alpha_j^- \exp\left({-i\int^{r}_{r_j} p(r') dr'}\right),
\end{multline}
where $\alpha_j^\pm$ are constants and we have omitted the radially dependent prefactor shared by both terms, as well as an overall factor of $\exp\left({im\theta-i\omega t}\right)$, which are irrelevant for our purposes. At $r_B$, we have the requirement that no fluid penetrates the outer boundary, 
\begin{equation}
    \partial_r\delta\Phi(r_B)=0, 
\end{equation}
which is solved by setting $\alpha_B^+=\alpha_B^-$. In the vicinity of $r_\mathrm{tp}$, it is possible to solve the wave equation~\eqref{eq:wave-eq} directly to find a relation between the WKB amplitudes. However, if the evanescent mode decays as it moves away from the turning point (manifestly apparent for state {I} in Fig.~\ref{fig:cobound}a) the relevant relation simply reads $\alpha^-_\mathrm{tp} = i\alpha^+_\mathrm{tp}$. Relating the amplitudes on either side of $r_\mathrm{tp}$, one obtains the resonance condition,
\begin{equation}
    \int^{r_B}_{r_\mathrm{tp}}p(r') dr' = \pi\left(n+\frac{1}{4}\right),
\end{equation}
where $n=0,1,2,\ldots$ indexes possible bound states that fit inside the cavity delineated by $r_\mathrm{tp}$ and $r_B$. We solve this condition to find the various $\omega_n$ and display the corresponding frequencies as open black circles in Fig.~\ref{fig:cobound}c.
\end{document}